\documentclass[12pt,twocolumn]{article}
\usepackage[a4paper,left=20mm,right=20mm,top=25mm,bottom=25mm,includeheadfoot]{geometry}

\usepackage{mathpazo}
\usepackage{graphicx}
\usepackage{amsmath,textcomp}
\usepackage{multirow}
\usepackage{makeidx}
\makeindex

\usepackage{sectsty}
	\allsectionsfont{\sffamily\raggedright}
	\sectionfont{\sffamily\large\raggedright}
	\subsectionfont{\sffamily\normalsize\raggedright}
\usepackage{ftnright}

\usepackage{widetext}
\usepackage{flushend}
\usepackage{cuted}
\usepackage{color}
\usepackage{graphicx}
\usepackage{hyperref}
\usepackage{pstricks,amssymb}

\usepackage{fancyhdr}
\begin{document}
\title{\vspace{-2em}\bfseries\sffamily Determining planetary positions in the sky for $\pm 50$ years to an accuracy of $\stackrel{<}{_{\sim}} 1^{\circ}$ with a calculator}
\author{\normalsize Tanmay Singal${^1}$ and 
Ashok K. Singal${^2}$\\[2ex]
$^{1}$Department of Physics and Center for Field Theory and Particle Physics,\\ Fudan University, Shanghai 200433, China.\\
{\tt tanmaysingal@gmail.com}\\[2ex]
$^{2}$Astronomy and Astrophysics Division, Physical Research Laboratory\\
Navrangpura, Ahmedabad 380 009, India.\\
{\tt ashokkumar.singal@gmail.com}
}
\date{\itshape Submitted on XX-XX-XXXX}
\maketitle
\thispagestyle{fancy}

\begin{abstract}
{\sffamily
In this paper, we describe a very simple method to calculate the positions of the planets in the
sky. The technique used enables us to calculate planetary positions to an accuracy of $\stackrel{<}{_{\sim}} 1^{\circ}$ for $\pm 50$ years from the starting epoch. Moreover, this involves very simple calculations 
and can be done using a calculator. All we need are the initial specifications of planetary orbits for some standard epoch and the time periods of their revolutions.
}\\ 
\hrule
\end{abstract}

\section{Introduction}
The night-sky fascinates people. To be able to locate a planet in the night sky is something 
that thrills people. Since the planets move with respect to the background stars and continuously change their positions in the sky, locating them in the sky, especially when seen from Earth that continuously shifts its position around Sun, could appear be a non-trivial 
task. It is a general notion that calculating the planetary positions is a very tedious task, 
involving a lot of complicated mathematical equations and computer work. However, to be able 
to locate planets in the sky one does not really need very accurate positions. After all, Kepler's 
laws, which describe planetary orbits reasonably well, are mathematically simple. Hence, one 
could use Kepler's law to predict planetary positions in which mutual influence of planets 
is not considered. Thereby an accuracy of $\stackrel{<}{_{\sim}} 1^{\circ}$ in planetary positions would be 
achieved.

In this paper, we employ a very simple method to calculate the positions of the planets. 
The technique we use enables us to calculate planetary positions to an accuracy of $\stackrel{<}{_{\sim}} 1^{\circ}$ 
for $\pm 50$ years from the starting epoch. Moreover, this involves very simple calculations 
and can be done using a calculator. All we need are the initial specifications of planetary orbits 
for some standard epoch and their time periods of revolution. Although accurate planetary positions
could be obtained easily from the internet, yet it is very instructive and much more satisfying
to be able to calculate these ourselves, starting from basic principles and using a simple procedure.

Our first step would be to calculate the positions of all the planets (including that of the Earth) in their 
orbits around the Sun. We initially consider the planets to revolve around the Sun in uniform 
circular motions. Knowing their original positions for the starting epoch, we calculate their 
approximate positions for the intended epoch.
As a consequence of this approximation there will be some errors since the actual orbits are 
elliptical and in an elliptical orbit the angular speed of the planet is not uniform and to an extent varies. Therefore, to get more accurate positions, we need to make appropriate corrections, which are derived in Appendix A. These corrections account for the elliptical motion.

Knowing the positions of the planets around the Sun, we can then use simple co-ordinate geometry to 
transform their position with respect to an observer on Earth. Our task becomes simple since the 
orbits of all planets more or less lie in the same plane, viz. the ecliptic plane.

Actually for locating planets in sky, even manual calculations could suffice, however, to  correct for the orbital ellipticity, numerical values of some trigonometric functions may be needed, otherwise one needs tabulated values of corrections [1]. 
Also, in order to transform the planet position to a geocentric perspective, a plot of the  relative positions of Sun, Earth and the planet on a graph sheet or a chart, using a scale and a protractor, may be required. However, both these tasks could be performed with the help of a scientific calculator. Further, if one wants to calculate Moon's position too (even though one could locate Moon in sky easily, without its position calculation), required for knowing the New Moon or the Full Moon dates, or the dates of possible solar or lunar eclipses in a specific year, computations could be done on a scientific calculator.

In this paper, we calculate the motion of naked-eye planets only, although the procedure can be 
applied equally well for the remaining planets also.
\begin{figure}[t]
\includegraphics[width=\linewidth]{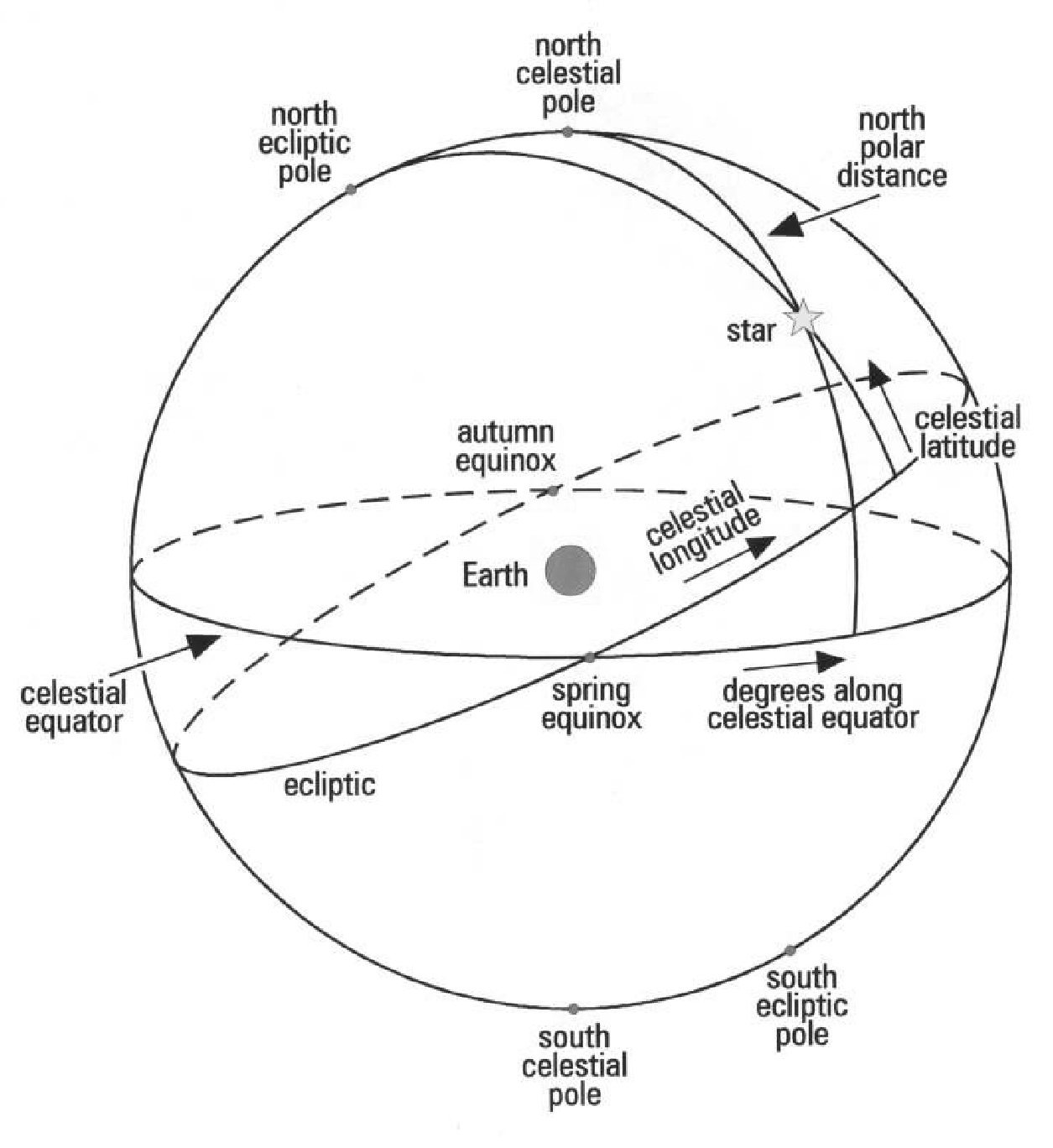}
\caption{Celestial sphere showing the ecliptic co-ordinate system [2].}
\end{figure}

\section{Celestial co-ordinates}

All celestial bodies in the sky, including stars, planets, Sun, Moon and other objects, appear to 
lie on the surface of a giant sphere called the Celestial Sphere. Due to Earth's eastward rotation 
around its axis, the celestial sphere appears to rotate westward around Earth in 24 hours. 
Infinitely extending the plane of Earth's equator into space it appears to intersect the celestial 
sphere to form a circle, which is called the Celestial Equator. 

As Earth moves around Sun - as seen from the Earth - Sun changes its position with respect to the 
background stars. The path that Sun takes on the celestial sphere is called the ``Ecliptic''. The 
familiar Zodiac constellations are just divisions of the ecliptic into twelve parts. Since all  
other planets revolve around Sun in nearly the same plane, they also appear to move on the ecliptic.

\begin{table*}[ht]
\begin{center}
\caption{Mean longitude $\lambda_{0}$ on 01/04/2020, 00:00 UT}
\vspace{-0.3cm}
\vbox{\columnwidth=33pc}
\begin{tabular}{lcccccc}
\hline
Planet  & $\lambda_{\rm i}$ ($^{\circ}$) &  $T$ (days)&  $\omega_{0}$ ($^{\circ}$/day) & $\lambda_{0}$ ($^{\circ}$) \\ 
\hline
Mercury & 250.2 & 87.969 & 4.09235 & 277.2  \\
Venus & 181.2 & 224.701 & 1.60213 & 150.6 \\
Earth & 100.0 & 365.256 & 0.98561  & 189.6  \\ 
Mars & 355.2 & 686.980 & 0.52403  & 270.9   \\
Jupiter & 34.3 & 4332.59 & 0.08309 & 288.8 \\
Saturn & 50.1 & 10759.2 & 0.03346  & 297.6  \\
\hline
\end{tabular}
\end{center}
\end{table*}

The celestial equator is inclined to the ecliptic by $23.5^{\circ}$. The points of intersections of these two 
circles on the celestial sphere are called the ``Vernal Equinox'' and the ``Autumnal Equinox''. The Vernal 
Equinox, also known as the Spring Equinox, is the point on the celestial sphere that the Sun passes through 
around 21st of March every year.

In astronomy, an epoch is a moment in time for which celestial co-ordinates or orbital elements  
are specified, while a celestial co-ordinate system is a co-ordinate system for mapping positions in the 
sky. There are different celestial co-ordinate systems, each using a co-ordinate grid  projected on the 
celestial sphere. The co-ordinate systems differ only in their choice of the fundamental plane, which 
divides the sky into two equal hemispheres  along a great circle . Each co-ordinate system is named for 
its choice of fundamental plane.

The ecliptic co-ordinate system is a celestial co-ordinate system that uses the ecliptic  for its 
fundamental plane. The longitudinal angle is called the ecliptic longitude or celestial longitude 
(denoted by $\lambda$), measured eastwards from $0^{\circ}$ to $360^{\circ}$ from the vernal equinox.
The latitudinal angle is called the ecliptic latitude or celestial latitude (denoted by $\beta$), 
measured positive towards the north. This coordinate system is particularly useful for charting 
solar system objects. 

The Earth's axis of rotation precesses around the ecliptic axis with a time period of about 25800 years. 
Due to this, the equinoxes shift westwards on the ecliptic. Due to the westward shift of the Vernal 
Equinox, which is the origin of the ecliptic co-ordinate system, the ecliptic longitude of the 
celestial bodies increases by an amount $360/258 \sim 1.4^{\circ}$ per century or a degree in $\sim 72$ years, the life span of a human.

Most planets, dwarf planets, and many small solar system bodies  have orbits 
with small inclinations to the ecliptic plane, and therefore their ecliptic latitude $\beta$ is 
always small. Due to only small deviations of the orbital planes of the planets from the plane of the ecliptic, the ecliptic longitude alone may suffice to locate planets in the sky.
\section{Calculating planetary positions}
\subsection{Heliocentric circular orbit}
Here, we consider the planets to move around Sun in circular orbits with a uniform angular speed.
The initial values of mean longitudes ($\lambda_{\rm i}$) of the planets given in Table 1 are 
for 1st of January, 2000 A.D., 00:00 UT (adapted from [3]). In Table 1, we have also listed the period, 
$T$ (days), of revolution of the planets [4]. Then, the mean angular speed is give by, 
$\omega_{0}=360/T$ ($^{\circ}$/day). We denote the Mean Longitude of the planets 
in the imaginary circular orbit for subsequent dates as $\lambda_{0}$.

We now demonstrate how to calculate $\lambda_{0}$ for Mars on April 1, 2020. 

The initial mean longitude, $\lambda_{\rm i}$, of Mars on 01.01.2000 at 00:00 UT  = 355.2$^{\circ}$.

Number of days between 01.01.2000 and 01.04.2020 = 7396.

Mean angle traversed duration this period $= 0.52403 \times 7396  = 3,875.7^{\circ}$.

So, on 01.04.2020 at 00:00 UT, the mean longitudes is

 $\lambda_{0}$,  $= 355.2 + 3,875.7  =  4,230.9 = 360 \times 11+270.9^{\circ}$.
 
We take out 11 integer number of complete orbits to obtain  $\lambda_{0}=270.9^{\circ}$.

In the same way, mean longitudes of all planets have been calculated in Table 1 for the same epoch.
\begin{table*}[h]
\begin{center}
\caption{Corrected longitude $\lambda$ on 01/04/2020, 00:00 UT}
\vspace{-0.3cm}
\vbox{\columnwidth=33pc}
\begin{tabular}{lccccccccc}
\hline
Planet & $\lambda_{0}$ ($^{\circ}$) &  $\lambda_{\rm p}$ ($^{\circ}$) & $\theta_{0}$ ($^{\circ}$) & $e$ 
& $\Delta\theta$ ($^{\circ}$) & $\theta$ ($^{\circ}$) &
$\lambda$ ($^{\circ}$) \\ 
\hline
Mercury & 277.2 & 77.5 & 199.7 & 0.2056 & -6.0  & 193.7 & 271.5\\
Venus & 150.6  &131.6 &  19.0 & 0.0068 & 0.3 & 19.2  & 151.1  \\
Earth & 189.6  & 102.9 & 86.7  & 0.0167 & 1.9  & 88.6  & 191.8 \\ 
Mars & 270.9 & 336.1 &  294.8 & 0.0934 & -10.2 & 284.6  & 261.0  \\
Jupiter & 288.8  &14.3 & 274.5 & 0.0485 & -5.6 & 269.0  & 283.6 \\
Saturn & 297.6   & 93.1 & 204.5 & 0.0555 & -2.5 & 202.0 & 295.4  \\
\hline
\end{tabular}
\end{center}
\end{table*}

\subsection{Heliocentric elliptical orbit}  
Our next step is to correct for the elliptical shape of the orbit as for some of the planets, depending upon the eccentricity ($e$), the corrections could be substantial. In a circular motion, the angular speed of the planet is constant, however, in an elliptical orbit, the angular speed of the planet varies with time. Due to this varying angular speed, actual longitude $\lambda$ of the planet will be somewhat different from the mean longitude $\lambda_{0}$.

Before we could make corrections for the elliptical shape of the orbit we need to know the orientation 
of the ellipse within the ecliptic and that can be defined by the longitude ($\lambda_{\rm p}$) of the perihelion. The distance of the planet from Sun varies in an elliptical orbit and 
perihelion is the point on the elliptical orbit that lies closest to Sun (the orbital point farthest from Sun is called aphelion). Longitudinal distance of the planet from the perihelion along the elliptical orbit is known as its ``anomaly'' (denoted by $\theta$), while angular distance of mean position of planet with 
respect to the perihelion is called the ``mean anomaly'' (denoted by $\theta_{0}$). As has been discussed 
in Appendix A, there is a one--to--one correspondence between $\theta$ and $\theta_{0}$.
\begin{table*}[h]
\begin{center}
\caption{Geocentric longitude $\lambda_{\rm g}$ on 01/04/2020, 00:00 UT}
\vspace{-0.3cm}
\vbox{\columnwidth=33pc}
\begin{tabular}{lccccccccc}
\hline
Planet & $a$ (A.U.) &  $e$ & $r$ (A.U.) & $\lambda$ ($^{\circ}$) & $r_{\rm g}$ (A.U.) & $\lambda_{\rm g}$ ($^{\circ}$)  \\ \hline
Mercury & 0.387  &  0.2056 & 0.463 & 271.5 &1.024 & 345.3 \\
Venus & 0.723  &0.0068 & 0.718 & 151.1 & 0.653 & 57.6 \\
Earth/Sun & 1.00  &0.0167 & 0.999 & 191.8  & 0.999 & 11.8  \\ 
Mars & 1.52 & 0.0934  & 1.472 &  261.0 &1.457 & 300.9  \\
Jupiter & 5.20 & 0.0485 & 5.192  & 283.6  &5.318 & 294.4  \\
Saturn & 9.55 & 0.0555  & 10.04 & 295.4 &10.32 & 300.8 \\
\hline
\end{tabular}
\end{center}
\end{table*}

The correction $\Delta\theta$ to be added to $\theta_{0}$ (Appendix A, Equation 4) is
\begin{eqnarray*}
\Delta\theta = 2\, e \sin \theta_{0} + \frac{5}{4}\, e^{2} \sin 2\theta_{0},
\end{eqnarray*}
where $e$ is the eccentricity of the ellipse.

In Table 2, we have listed values of the longitude of perihelion ($\lambda_{\rm p}$) and eccentricity ($e$) for all planets, taken from [3]. 

Let's consider Mercury's position on 01/04/20 at 00:00 UT.

Mean longitude, $\lambda_{0}$ = 277.2$^{\circ}$

Perihelion Longitude, $\lambda_{\rm p} = 77.5^{\circ}$
				
Mean anomaly, $\theta_{0} =\lambda_{0} - \lambda_{\rm p} = 199.7^{\circ}$

The 1st order correction then is 

$\Delta\theta_1=2 e \sin \theta_{0}= 2 \times 0.2056 \times \sin (199.7^{\circ}) 
= -0.13861$ rad $= -7.9^{\circ}$,

while the 2nd order correction is

 $\Delta\theta_2= \frac{5}{4} e^{2} \sin 2\theta_{0}
 = 1.25 \times (0.2056)^2 \times  \sin  (39.4^{\circ})  = 0.03354$ rad $= 1.9^{\circ}$.

The total correction then is

$\Delta\theta$ = $\Delta\theta_1 + \Delta\theta_2 = -7.9 +1.9 = -6.0^{\circ}$

The anomaly, $\theta = \theta_{0} +  \Delta\theta = 199.7 -6.0 = 193.7^{\circ}$

Precession of vernal equinox in 20.25 years $= 20.25 \times 360/25800  \approx 0.3^{\circ}$.

The longitude, $\lambda = \lambda_{0} + \Delta\theta $ + precession of vernal equinox 
$=  277.2 -6.0 + 0.3 = 271.5^{\circ}$.

We can obtain corrections for the elliptical orbits of the remaining planets in the same way. 
The calculated anomaly ($\theta$) and 
longitude ($\lambda$) are listed in Table 2. The residual errors in $\lambda$ values are $\stackrel{<}{_{\sim}} 1^{\circ}$. 

\subsection{Geocentric perspective}
Until now, we have calculated the longitudes, $\lambda$, of the planets on the celestial sphere centred on 
the Sun. We can also calculate radii $r$ of their orbits around the sun, for that epoch, giving their positions in 
polar form ($r, \lambda$). To get the positions of planets on the celestial sphere centred on the Earth, 
we first convert the polar co-ordinates into rectangular form with origin on Sun, and after shifting the origin from Sun to Earth, we change them back into polar form. 

For converting into a rectangular form, we have to decide upon the direction of the X and Y axes. 
We assume X to be in the positive direction along the line joining Sun to the Vernal Equinox, and
Y to be perpendicular to X in the ecliptic plane in such a way that the longitude is a positive angle. 

\subsection{An example}
As an example, this procedure is demonstrated for Mercury's position on 01/04/20 at 00:00 UT.
\subsubsection{Heliocentric co-ordinates} 
Distance, $r$, of Mercury from Sun can be obtained from the anomaly $\theta$ as,
\begin{eqnarray*}
 r = \frac {a(1-e^2)} {1 + e \cos \theta}  = 0.463 {\rm A.U.},
\end{eqnarray*}
where $ a = 0.387$ A.U. is the length of semi--major axis of its elliptical orbit. Distances to planets in the Solar system are conventionally measured in astronomical units (A.U.), the mean distance of Earth from Sun, which is nearly $1.5 \times 10^8$ km. 

The heliocentric polar co-ordinates of Mercury thus are (0.463 A.U., 271.5$^{\circ}$).  Then we can
get heliocentric rectangular co-ordinates of Mercury as,

$X_{\rm h} = r \cos(\lambda) = 0.012$ A.U.

$Y_{\rm h} = r \sin(\lambda) = -0.463$ A.U.

Similarly we get heliocentric rectangular co-ordinates of Earth as, 

$X_{0} = -0.978$ A.U.

$Y_{0} = -0.204$ A.U.

\subsubsection{Geocentric co-ordinates} 
Geocentric rectangular co-ordinates of Mercury then are

$X_{\rm g} = X_{\rm h} - X_{0} = 0.990$ A.U.

$Y_{\rm g} = Y_{\rm h} - Y_{0} = -0.259$ A.U.

Converting these into polar form, we get the geocentric distance and longitude as,

$r_{\rm g}= \sqrt {(X_{\rm g}^2+Y_{\rm g}^2)} = 1.024$ A.U.

$\lambda_{\rm g} = \tan ^{-1} (Y_{\rm g}/X_{\rm g}) = 345.3^{\circ}$.

In Table~3 we have given the calculated geocentric longitudes of various planets on 01.04.2020 at 00:00 UT. 
In the Earth/Sun row in Table~3, $r_{\rm g}$, $\lambda_{\rm g}$  
are the geocentric values for the Sun's position. 
The position of Sun on the celestial 
sphere, as seen from Earth, is in a direction exactly opposite to that of Earth as seen from the Sun. 
Therefore the geocentric longitude of Sun is the heliocentric longitude of Earth plus 180$^{\circ}$.  

We have ignored any perturbations on the motion of a planet due to the effect of other planets which may 
distort its elliptical path. We are able to get the accuracy of $\stackrel{<}{_{\sim}} 1^{\circ}$ for long periods ($\pm 50$ years) 
because most of the terms ignored in the heliocentric longitude calculations are periodic in nature and do 
not grow indefinitely with time (see e.g. [5]). The other parameters characterizing the elliptical 
orbit, like the longitude of the perihelion, semi--major axis and eccentricity etc. change so slowly with time 
that for the accuracy we are interested in, these can be considered constant for $\pm 50$ years.

\section{Locating planets in sky}

After having computed the geocentric longitudes of the planets, we are now in a position to locate them 
in the sky. Any one familiar with the Zodiac constellations could locate a planet from its position in the
constellation in which it lies. The ecliptic is divided into 12 Zodiac signs -- Aries, Taurus, Gemini, Cancer, 
Leo, Virgo, Libra, Scorpio, Sagittarius, Capricorn, Aquarius, Pisces. The Vernal equinox, at zero ecliptic longitude, 
is the start of the first Zodiac sign and is also known as the First Point of Aries. But there is a caveat attached. 
Because of the precession the vernal equinox has shifted westward by almost the full width of a constellation in the 
last $\sim 2000$ years since when the Zodiac signs and constellation were perhaps first identified. As a consequence, 
the First Point of Aries now lies in the constellation Pisces.  For example, on 01/04/2020, geocentric longitude 
$294.4^{\circ}$ of Jupiter implies it is in the 10th Zodiac sign Capricorn, but actually lies in the Sagittarius 
constellation, taking into account the shift by one constellation due to precession. There are further complications. 
The twelve constellations are not all of equal length of arc along the ecliptic longitude. Moreover there is another 
constellation, viz. Ophiuchus, through which the ecliptic passes. However these complications are somewhat set 
aside by the fact that there are only about half a dozen stars in the Zodiac with an apparent brilliance comparable 
to the naked--eye planets, therefore with some familiarity of the night-sky, one could locate the planets easily 
from their geocentric longitude values. It further helps to remember that unlike stars, the planets, because of their 
large angular sizes, do not twinkle.

For a more precise location of a planet we can calculate its relative angular distance from the Sun along the ecliptic. 
The difference between the geocentric positions of a planet and Sun (Table 4) is called the 
elongation ($\psi$) of the planet and it tells us about planet's position in the sky with respect to that of the Sun. The longitude increases eastwards, therefore, if the longitude of the planet is greater than that of the Sun, then the planet lies to the east of the Sun. That means, in the morning 
the Sun will be rising before the planet rises but in the evening the planet will be setting after the sunset. 
So the planet will be visible in the evening sky in the west. On the other hand, if the geocentric 
longitude of the planet is smaller than that of the Sun, then it lies to the west of the Sun and  will rise before the sunrise and will be visible in the morning in the eastern sky.

\begin{table}[t]
\begin{center}
\caption{Elongations of the planets on 01/04/2020 
}
\vspace{-0.8cm}
\vbox{\columnwidth=33pc}
\begin{tabular}{lcccccc}
\hline
\multirow{2} {*}{Planet}  & $\lambda_{\rm g}(^{\circ})$ & $\psi(^{\circ})$ & &  $\lambda_{\rm g}(^{\circ})$ & $\psi(^{\circ})$   \\ 
 \cline{2-3} \cline{5-6}
& \multicolumn{2}{c}{05:30 IST} & & \multicolumn{2}{c}{(17:30 IST)} \\           
\hline
Sun & 11.8  & - & &12.3 & - & \\ 
Mercury & 345.3 & -26.4 & &346.0& -26.3   \\
Venus & 57.6  & 45.9 & &58.1& 45.8\\
Mars &  300.9 & -70.9 & &301.3 & -71.0  \\
Jupiter & 294.4 & -77.4 & &294.4 & -77.8  \\
Saturn & 300.8 & -71.0 & &300.8& -71.4  \\
\hline
\end{tabular}
\end{center}
\end{table}

\begin{figure}[t]
\includegraphics[width=\linewidth]{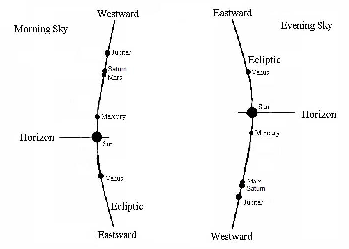}
\caption{A schematic representation of the elongations of planets at the times of sunrise
and sunset on April 1, 2020.
}
\end{figure}

In Table 4, positions of planets with respect to Sun on 01.04.2020 are given for 00:00 UT,
which corresponds to 05:30 IST (Indian Standard Time). 
For example, the geocentric longitude of Venus with respect to Sun is $57.6-11.8 = 45.9^{\circ}$. Thus Venus has an eastern elongation $45.9^{\circ}$ on 01.04.2020, 05:30 IST, and would be visible only in the evening, after the sunset, in the western sky.

As Earth completes a rotation in 24 hours, the westward motion of the sky is at a rate $360/24=15^{\circ}$/ hour.
This rate is strictly true for the celestial equator, but we can use this as an approximate rotation rate 
even for the ecliptic, which is inclined at $23.5^{\circ}$ to the equator. Therefore Mercury,    on 01.04.2020, will rise about $26.4/15\sim 1.5$ hours before the sunrise, and could be seen in the eastern sky in the morning.

It is, however, possible to determine the planetary positions for any other time of the day. For instance, in the case of Venus, which is visible in the evening hours on our chosen date of 1/4/2020, it might be preferable to calculate the position on that date for 12:00 UT, corresponding to 17:30 hr IST, locally an evening time. For this, we use the number of days in Step 1 as 7396.5.  It should be noted that not only the longitude of each planet around Sun might change by a certain amount, even the longitude of Earth advances by $\sim 1^{\circ}$ in a day, thus affecting the elongation of even Jupiter and Saturn (Table 4), whose angular speeds are relatively small (Table 1).
In Table 4, we have listed the elongations of all five naked-eye planets on 01-04-2020 at 00:00 hr UT (5:30 IST) and at 12:00 hr UT (17:30 IST). 
Venus with an eastern elongation 
$\sim 46^{\circ}$ on that evening, will be setting approximately three hours after the sunset. 
This way, one can easily locate the planets in the sky from their elongations.

Figure 2 is a schematic representation of the relative positions of various planets with respect to the Sun at the horizon, on the morning and evening of April 1, 2020.
\section{Conclusions}

It is generally thought that calculation of position of planets in the night sky is a difficult job, 
which can be accomplished only by complex scientific computations, using fast computers. Our motive 
here has been to dispel such a notion and bring out the fact that such complex and accurate computations are not 
always really necessary. One can calculate the position of planets using the method derived here and 
get the thrill of finding the planet at the predicted position in the night sky.

We have been able to obtain the position of planets within an accuracy of $\stackrel{<}{_{\sim}} 1^{\circ}$., using a calculator. 
This method can be used to reckon planetary positions up to $\pm 50$ years of the starting epoch.

\section*{Appendix A: Correction for the orbital ellipticity} 
We compute the correction for motion of a planet in an actual elliptical orbit from that in an 
imaginary circular orbit.
Period of revolution in circular orbit is taken to be exactly the same as that in the elliptical orbit. 
The origin of the mean longitude in circular orbit is chosen such that $\lambda_{0}$ coincides with 
the longitude $\lambda$ of the planet when it is at the perihelion in its elliptical orbit. 
\begin{figure*}[ht]
\includegraphics[width=\linewidth]{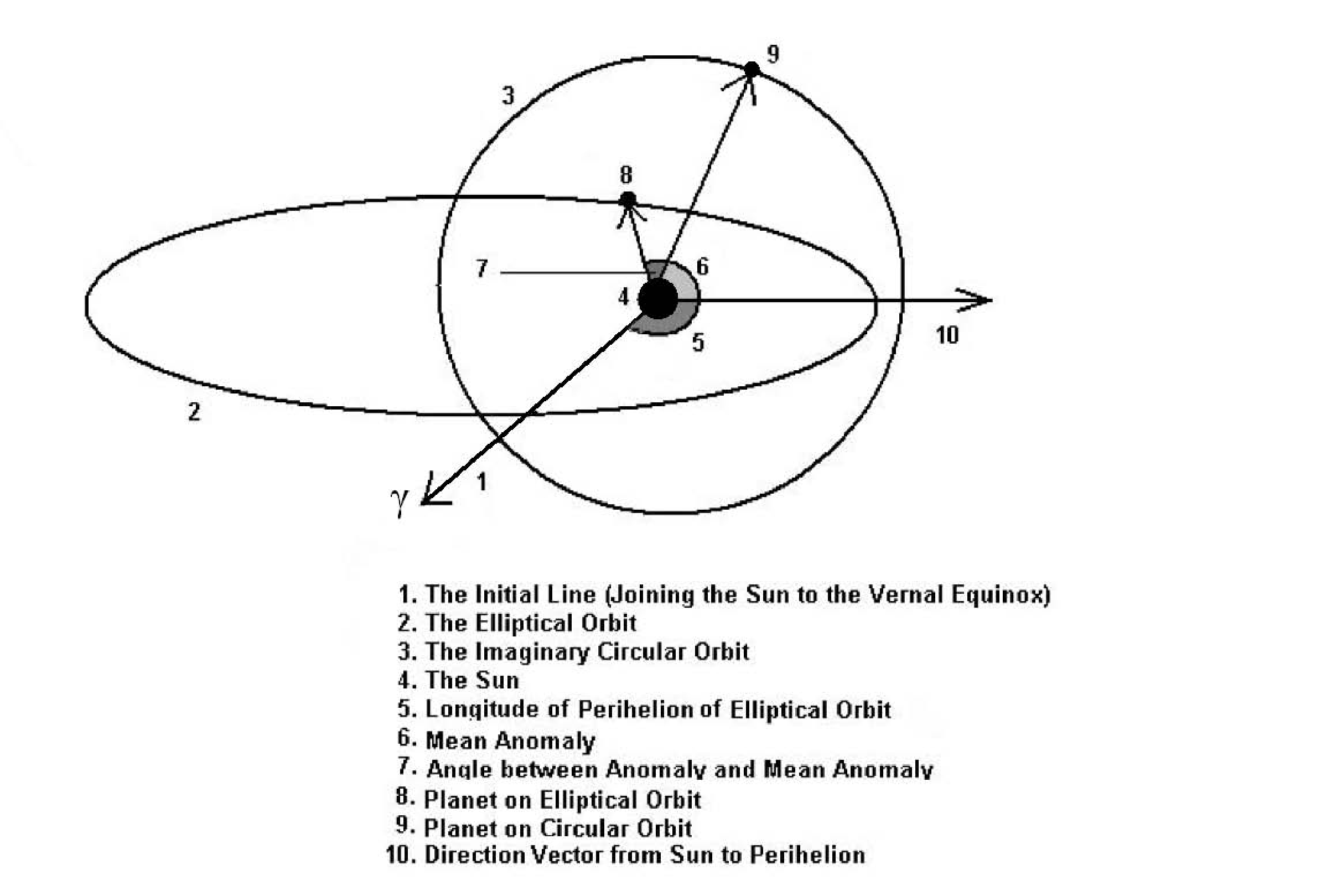}
\caption{A schematic diagram of the planet in circular and elliptical motion.}
\end{figure*}
For mathematical convenience, we take $t=0$ at that instant. Then $\lambda_{0}(0) = \lambda(0)$.
Let $\lambda_{\rm p}$ be the longitude of the perihelion of the planet's elliptical orbit. We subtract 
$\lambda_{\rm p}$  from $\lambda_{0}$ and $\lambda$  to obtain what is called respectively, the mean anomaly $\theta_{0}$ and the anomaly $\theta$ of the planet. Thus 
 $\theta_{0}$ = $\lambda_{0}$ - $\lambda_{\rm p}$, and 
 $\theta$ = $\lambda$  - $\lambda_{\rm p}$.

Then $\theta_{0}(0) = \theta(0)$

In circular motion, the angular speed, $\omega_{0}$, of the planet is constant. However, in the elliptical motion, 
the angular speed of the planet varies with time. 

Let a time $t$ has passed after $t=0$. Then, the change in $\theta_{0}$ of the planet is 
$\omega_{0} t$ whereas the change in $\theta$ of the planet won't be the same because of the variation in 
the angular speed along elliptical trajectory. 

Let $\Delta \theta(t)= \theta(t)$ -  $\theta_{0}(t)$.

We know that $\theta_{0}(t)$ and $\theta(t)$ are periodic by the same time 
interval, $T$, as $T$ is the time period of revolution in both the cases (elliptical and circular 
motion). Hence, all value of $\theta_{0}(t)$ and $\theta(t)$ repeat after a time period of T. 
Hence, $\theta_{0}(t)$ and $\theta(t)$  have a 
one--to--one relation. Hence, $\Delta \theta$ also repeats after time $T$. The uniform circular motion 
thus is a useful approximation because the error $\Delta \theta$ is periodic with time and does 
not keep accumulating with time to grow to very large values.

To find the correction, first consider an elliptical orbit of a planet around the Sun as shown in Figure 3. 
We use the equation of the ellipse in polar co-ordinates ($r, \theta$) where $\theta$ is the anomaly. 
The equation of the ellipse then is
\begin{eqnarray}
r = \frac {l}{1 + e \cos \theta} = \frac {a (1-e^2)}{1 + e \cos \theta} ,
\end{eqnarray}
where $l = a(1-e^2)$ is the semi--latus rectum with $a$ as the semi--major axis and $e$ the eccentricity of the ellipse. 
The semi--minor axis of the ellipse is $b = a \sqrt{1-e^2}$. 

Now, total area of the ellipse $ A = \pi a b$ is swept in $T$, the time period of revolution. From 
Kepler's second law we know that the rate of area swept out by the position vector of planet (with respect to Sun) 
is a constant. Therefore the rate of area swept is
\begin{eqnarray*}
\frac {{\rm d} A}{{\rm d} t} = \frac {r^2}{2}  \frac {{\rm d} \theta}{{\rm d} t} = \frac {\pi a b}{T}.
\end{eqnarray*}
Substituting from Equation (1), we get
\begin{eqnarray*}
\frac {2 \pi}{T} = \frac {(1-e^2)^{\frac {3}{2}}} {(1 + e \cos \theta)^{2}}\, \frac {{\rm d} \theta}{{\rm d} t}. 
\end{eqnarray*}
We notice that ${2 \pi}/{T}$ is nothing but the mean angular speed $\omega_{0}$. Therefore
\begin{eqnarray}
\theta_{0}(t)=\int_{0}^{t}\omega_{0}\, {\rm d t} = \int_{0}^{t}\frac{(1-e^2)^{\frac {3}{2}}} 
{(1 + e \cos \theta)^{2}}\,\, {\rm d} \theta.
\end{eqnarray}
We want to get the equation in the form, $\theta=\theta_{0}+\Delta\theta$, 
so that by adding the longitude of the perihelion on both sides of the equation, we could get the 
relation between the correct longitude $\lambda$ and the mean longitude $\lambda_{0}$. 

A direct integration of Equation (2) may not be possible, but we can expand the integrand as a series and 
integrate only a few first most significant terms.  A binomial series expansion is possible because the 
eccentricity of an ellipse, $e < 1$. During the expansion we drop terms having higher than  $e^{2}$ factor.
\begin{eqnarray*}
\theta_{0}(t)= \int_{0}^{t}(1-\frac {3}{2}\,e^2+\cdots)(1 -2\,e \cos \theta \\
+ 3\,e^2 \cos^2 \theta+\cdots)\, {\rm d} \theta 
\end{eqnarray*}
After integration we get
\begin{eqnarray}
\theta_{0}= \theta - 2\, e \sin \theta + \frac{3}{4}\, e^{2} \sin 2\theta + \cdots,
\end{eqnarray}

which can be written as 
\begin{eqnarray*}
\Delta \theta = \theta-\theta_{0}=  2\, e \sin \theta - \frac{3}{4}\, e^{2} \sin 2\theta \\
+ \cdots
\end{eqnarray*}
However we want the right hand side of Equation (3) to be expressed in terms of $\theta_{0}$. For that we can substitute 
$ \theta = \theta_{0}+\Delta \theta$ on the right hand side to get,
\begin{eqnarray*}
\Delta \theta = 2\, e \sin (\theta_{0}+\Delta \theta) - \frac{3}{4}\, e^{2} \sin [2(\theta_{0}+\Delta \theta)]\\
+ \cdots
\end{eqnarray*}
Expanding in powers of $\Delta \theta$ and neglecting terms of order $e \,(\Delta \theta)^2$, $e^2 \,\Delta \theta$
 and higher we get
\begin{eqnarray*}
\Delta \theta \,(1-2\, e \cos \theta_{0}) = (2\, e \sin \theta_{0} - \frac{3}{4}\, e^{2} \sin 2\theta_{0}),
\end{eqnarray*}
or 
\begin{eqnarray*}
\Delta \theta = \frac{(2\, e \sin \theta_{0} - \frac{3}{4}\, e^{2} \sin 2\theta_{0})}
{(1-2\, e \cos \theta_{0})}.
\end{eqnarray*}
Again Expanding in powers of $e$ and ignoring terms of order $e^{3}$ or higher, we get
\begin{eqnarray}
\Delta \theta =  2 \,e \sin \theta_{0} + \frac{5}{4}\, e^{2} \sin 2\theta_{0} 
\end{eqnarray}
which is the required expression for the correction term.

\section*{Appendix B: Position of the Moon}
Here we determine Moon's position for any given epoch, say, April 1, 2020, 00 UT, starting from the initial epoch 1st January  2000, 00 UT. 

Moon moves in a {\em geocentric} orbit of mean eccentricity $e=0.0549$ and a tropical revolution period of 
$T=27.32158$ days, corresponding to a mean angular speed $\omega_{0}=13.17640^{\circ}$/day with respect to the 
vernal equinox. From the initial value $\lambda_{\rm gi}= 211.7^{\circ}$ on 1st of January, 2000 A.D., 00:00 UT, Moon's mean geocentric longitude is calculated as,
$\lambda_{\rm g0}=211.7 + 13.17640 \times 7396= 104.4^{\circ}$. 

But before we correct for the ellipticity of the Moon's orbit, we need to consider that unlike in a planetary 
orbit where the position of the perihelion change so slowly with time that it can be considered constant for 
$\pm 50$ years, in Moon's orbit the perigee rotates forward with respect to the vernal equinox with a period 
of 3231.4 days ($\sim 8.85$ years), corresponding to an angular speed $0.11141 ^{\circ}$/day. With an initial 
value of $ 83.3^{\circ}$, the longitude of the perigee, on April 1, 2020, 00:00 UT is then given by,
$\lambda_{\rm p}=83.3 + 0.11141 \times 7396= 187.3^{\circ}$. 
From this we get Moon's mean anomaly for a circular motion as, 
$\theta_{0}=\lambda_{\rm g0}-\lambda_{\rm p}=277.1^{\circ}$.

We can now use Equation (4) in the same way as for the planets to get the correction for the ellipticity 
$\Delta\theta = -6.3^{\circ}$, which gives Moon's corrected geocentric longitude $\lambda_{\rm g}=\lambda_{\rm g0}+\Delta\theta =98.1^{\circ}$. However, the value obtained 
thus is accurate only up to $\sim 2^{\circ}$. The reason being that there is an important perturbation term, known as 
Evection, which depends upon Moon's mean elongation $\psi_{0}$ as well as its mean anomaly $\theta_{0}$, and has a value, 
$\Delta\theta_{\rm ev} = 1.27^{\circ} \sin (2\psi_{0}-\theta_{0})$.
Substituting for $\psi_{0}=104.4-9.6=94.8^{\circ}$, we get, 
 $\Delta\theta_{\rm ev} = 1.27 \sin (2\times 94.8 -277.1)= -1.3^{\circ}$.
Thus a more accurate value of Moon's geocentric longitude on April 1, 2020, 00:00 UT is $\lambda_{\rm g}=98.1-1.3=96.8^{\circ}$, and the elongation $\psi$ 
calculated thence is $85.0^{\circ}$. 

There is another feature of the Moon's motion which has a direct bearing on the time of occurrences of solar and 
lunar eclipses. Moon's orbit is inclined to the ecliptic with a mean inclination of $5.16^{\circ}$, intersecting 
it at points known as the ascending and descending nodes. The ascending node is the one where Moon crosses the 
ecliptic in a northward direction. Gravitational pull of the Sun causes a precession of the axis of Moon's orbital 
plane around that of the ecliptic with a tropical period of 6798.4 days ($\sim 18.6$ years). Consequently the 
nodes of the Moon's orbit have a retrograde motion of $\sim 0.05295^{\circ}$/day around the ecliptic. 
With an initial value of $125.1^{\circ}$ on 1st January  2000, 00 UT, the longitude of the ascending node 
for the epoch April 1, 2020, 00:00 UT is  
$125.1-0.05295 \times 7396 = -266.5^{\circ}$, implying $\psi_\Omega=81.7^{\circ}$,
while the descending node is $180^{\circ}$ away at $\psi_\mho=-98.3^{\circ}$.

A solar eclipse can take place near the New Moon time, when Moon's elongation is $\sim 0^{\circ}$ and it could block sunlight to reach some parts of the Earth's surface, while a lunar eclipse can take place around the Full Moon time, when Moon's elongation is $\sim 180^{\circ}$ and Earth may be blocking sunlight from reaching the Moon. However, these events could occur only when Sun's longitude is close to that of one of the nodes because it is then only Earth and Moon both get nearly aligned in a straight line with Sun, so that one of these celestial bodies could block sunlight to cause a shadow on the other celestial body.

\section*{Appendix C: The astronomical calendar}
In all the examples presented above, the calculations were made using a scientific calculator. This procedure is appropriate 
for quickly getting planetary positions, to locate one or more planets in sky, occasionally, on some specific day. However if one wants to do 
many computations, say calculate positions for planets for many more days of the year, the process could become tedious and the 
chances of a numerical mistake occurring in manual calculations could become high. Since the process of computing planetary positions 
is a repetitive one, it could then be much more convenient to write a simple computer programme, using the algorithm described above, to carry out calculations. We have written such a programme to compute positions of all the planets as well as of the 
Moon for each day of a specified year and present the results in the form of an astronomical calendar which gives elongations of 
different planets for all days of the year. 
\begin{figure*}[t]
\includegraphics[width=\linewidth]{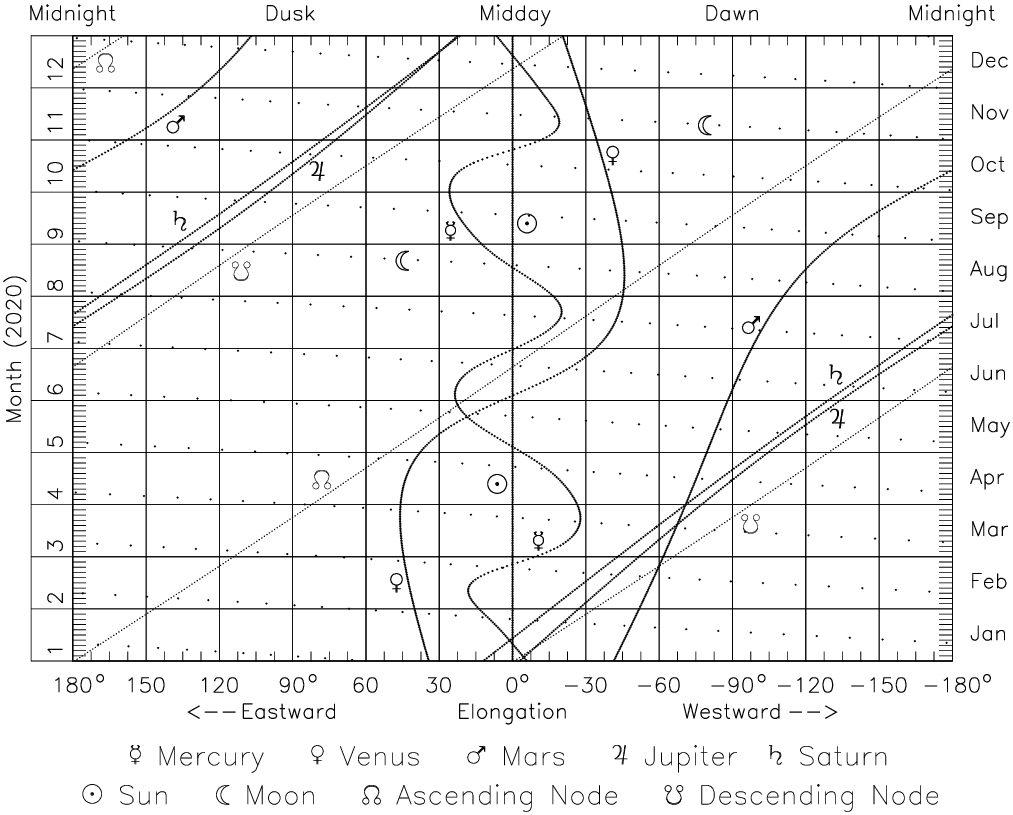}
\caption{Astronomical calendar for locating planets in sky for the year 2020.
}
\end{figure*}

The calendar allows us to locate the  naked-eye planets in the sky for any time of the corresponding year. 
The horizontal axis displays the elongation in 
degrees,
and is centred 
around the Sun, which by definition has a zero elongation. The vertical axis marks the day of the year. 
Thus to locate a planet on any given date of the year, we select that date on the vertical axis and then move 
in a horizontal direction till we find the planet. From the elongation of the planet we can easily locate it in the sky. Different paths of various planets as well as those of Sun, Moon and both nodes, can be identified from their conventional symbols, which are explained at the bottom of the figure.
We can use the calendar to find what all planets are above the horizon at any time of the day. Suppose for a given 
date of the year we want to locate all planets visible in the sky at, say, dawn. The Sun, at $0^\circ$ elongation, will 
at that time be just rising near the eastern horizon and the $-180^\circ$ elongation point in the calendar will be near 
the western horizon. The intermediate elongation points will be at in-between positions on the celestial hemisphere, 
e.g., the $-90^\circ$ elongation point will be close to the culmination point (the point nearest to the zenith). This way 
going along the horizontal direction from $0^\circ$ to $-180^\circ$ at the chosen date, we will find 
the celestial position of all planets visible in the morning sky on that date. At dusk, with sun setting near the 
western horizon, the visible sky will stretch eastward from $0^\circ$ to $180^\circ$ elongation on the calendar. Similarly 
one can locate planets on the celestial sphere at other hours of the day. At midnight, with culmination point 
being at $180^\circ$ (which is the same as $-180^\circ$), the sky towards west will stretch from $180^\circ$ to $90^\circ$ and that
towards east will be from $-180^\circ$ to $-90^\circ$ elongation, while at 9 p.m., with the culmination point at $135^\circ$, 
the celestial hemisphere will stretch from west to east between $45^\circ$ and $225^\circ$ (equivalently  $-135^\circ$) elongations.

The slant dotted lines running across the calendar from west to east almost every month represent Moon's path.
Their dates of intersection with the Sun's path at $0^\circ$ elongation mean New Moon days, while intersections with the 
midnight lines (at $\pm 180^\circ$ elongation) imply Full Moon days, with the intermediate phases at the in-between dates. 
The faint lines in the calendar represent the relative positions of the ascending node  and descending node of Moon's orbital plane. 
Intersection of the sun's path (at $0^\circ$ elongation) or of the midnight lines (at $\pm 180^\circ$) with one of the lines of nodes, 
indicate the possibility of occurrence of an eclipse. In the neighbourhood of these intersection points, at the time of a New 
moon there might possibly occur a solar eclipse while at the Full moon time there is a possibility of lunar eclipse.

For many planetary phenomena, the astronomical calendar can act as a quick indicator. However, because of the limited resolution of the display in the calendar, for a better accuracy one might go back to the actual 
tabulated data from which the calendar has been generated. From our computed data for the longitudes of the Sun, Moon and the lines of 
nodes for the year 2020, we find that the possible dates for the solar eclipses in the year 2020 are 21st June and 14th December, which are the New Moon date close to the intersection point of the lines of nodes with Midday line (Sun's path at $ 0^\circ$).
Similarly, possible lunar eclipses on Full Moon dates near 
the intersection points of the lines of nodes with Midnight lines (at $\pm 180^\circ$) would in 2020 fall on January 10, June 5, July 5 and November 30. However, as seen in Figure 4, these might only be Penumbral Lunar eclipses as the Full Moon positions do not fall very close to any of the intersection points.

From the astronomical calendar (Figure 4), we can expect some interesting astronomical vision in the year 2020. Jupiter and Saturn will be moving closer to each other in sky with the two being closest (within a degree  -- the accuracy limits of our calculations), an event known as a great conjunction, around 23rd of December. For about 10 days, near the end of March, when Jupiter and Saturn will be within $\sim 7^\circ$ of each other, Mars will be lying in between the two. Not only will there be three bright objects seen in close proximity of each other in sky, Mars, seen within a degree of Jupiter on 22nd of March will move to within a degree of Saturn by 1st of April.
But this spectacular sight will be for the eyes of the early birds only, as this would be visible before sunrise, in the eastern sky.

\end{document}